\documentclass{emulateapj}

\usepackage{natbib}
\usepackage{amsmath}

\shorttitle{TWO NEAR-INFRARED STATES OF SGR A*: FLARES AND LOW-LEVEL VARIABILITY}
\shortauthors{DODDS-EDEN ET AL.}

\begin{document}

\title{The two states of Sgr~A* in the near-infrared: \\bright episodic flares on top of low-level continuous variability
}
\author{K.~Dodds-Eden\altaffilmark{1}, S.~Gillessen\altaffilmark{1}, T.K.~Fritz\altaffilmark{1}, F.~Eisenhauer\altaffilmark{1},  S.~Trippe\altaffilmark{2}, R.~Genzel\altaffilmark{1,3},
T.~Ott\altaffilmark{1}, 
H.~Bartko\altaffilmark{1},  O.~Pfuhl\altaffilmark{1}, G.~Bower\altaffilmark{4}, A.~Goldwurm\altaffilmark{5,6}, D.~Porquet\altaffilmark{7}, G.~Trap\altaffilmark{5,6}, F.~Yusef-Zadeh\altaffilmark{8}}
\altaffiltext{1}{Max Planck Institut f{\"u}r Extraterrestrische Physik, Postfach 1312, D-85741, Garching, Germany.}
\altaffiltext{2}{Institut de Radioastronomie Millimetrique (IRAM), 300 rue de la Piscine,
38406 Saint Martin d'Heres,
France.}
\altaffiltext{3}{Department of Physics, University of California, Berkeley, 366 Le Comte Hall, Berkeley, CA 94720-7300}
\altaffiltext{4}{Department of Astronomy, University of California, Berkeley, 601 Campbell Hall, Berkeley, CA 94720-3411.}
\altaffiltext{5}{CEA, IRFU, Service d'Astrophysique, Centre de Saclay, F-91191 Gif-surYvette, France.}
\altaffiltext{6}{AstroParticule et Cosmologie (APC), 10 rue Alice Domont et Leonie Duquet, F-75205 Paris, France.}
\altaffiltext{7}{Observatoire astronomique de Strasbourg, Universit{\'e} de Strasbourg, CNRS, INSU, 11 rue de l'Universit{\'e}, F-67000 Strasbourg, France}
\altaffiltext{8}{Department of Physics and Astronomy, Northwestern University, Evanston, Il. 60208}


\begin{abstract}
In this paper we examine properties of the variable source Sgr~A* in the near-infrared (NIR) using a very extensive Ks-band data set from NACO/VLT observations taken 2004 to 2009. We investigate the variability of Sgr~A* with two different photometric methods and analyze its flux distribution. We find Sgr~A* is continuously emitting and continuously variable in the near-infrared, with some variability occurring on timescales as long as weeks. The flux distribution can be described by a lognormal distribution 
at low intrinsic fluxes ($\lesssim5$~mJy, dereddened with $A_{\rm Ks}=2.5$). The lognormal distribution has a median flux of $\approx$1.1~mJy, but above 5~mJy the flux distribution is significantly flatter (high flux events are more common) than expected for the extrapolation of the lognormal distribution to high fluxes. We make a general identification of the low level emission above 5~mJy as \emph{flaring} emission and of the low level emission as the \emph{quiescent state}. 
We also report here the brightest Ks-band flare ever observed (from August~5th, 2008) which reached an intrinsic Ks-band flux of 27.5~mJy ($m_{\rm Ks}=13.5$). This flare was a factor 27 increase over the median flux of Sgr~A*, close to double the brightness of the star S2, and 40\% brighter than the next brightest flare ever observed from Sgr~A*. 
\end{abstract}

\maketitle

\section{Introduction}
 The \emph{very} center of our galaxy houses the variable source named Sgr~A*, first discovered in the radio as a compact source \citep{bal1974}. The fact that this source is motionless (to better than 1 km/s) at the dynamical center of the galaxy \citep{Reid2004}, and its coincidence with the common focus of elliptical orbits of stars tracked over the last decade and a half, clearly associates it with a supermassive black hole of $4\times 10^6 M_\odot$ \citep{Schoedel2002,Ghez2008,Gillessen2009}. 

Sgr~A* has been detected across the electromagnetic spectrum, at radio, submm, NIR and X-ray wavelengths.
At NIR and X-ray wavelengths \citep{Genzel2003,Baganoff2001} the emission is highly variable (factors up to $\approx$160 and 27 in the X-ray and NIR respectively; \citealt{por08}, this work) compared to the comparatively steady emission at longer wavelengths. 
NIR peaks are detected more often than in the X-ray (peaks occur $\approx$1 and 4 times a day for X-ray and NIR variable emission, respectively; \citealt{Baganoff2003HEAD,Eckart2006_FlareActivity}). Some NIR flares have been detected without any accompanying X-ray flare \citep{Hornstein2007}. However, when both NIR and X-ray exhibit increases in emission, the peaks in emission occur simultaneously \citep[e.g.,][]{Eckart2004,DoddsEden2009}.

The near-infrared lightcurves from Sgr~A* exhibit $\sim$1~hour long increases in emission that are often called `flares' in the literature. A number of these have exhibited very suggestive substructural features with timescales of $\sim20$~minutes  \citep[][]{Genzel2003,eck06,Trippe2007,Eckart2008_NIRpol,Eckart2008_NIRsubmm,DoddsEden2009}, possibly quasi-periodic oscillations (QPOs). However the existence of QPOs and even the use of the term flare to describe the NIR variability of Sgr~A* has been questioned by \citet{Do2009} (see also \citealt{Meyer2008} and \citealt{Meyer2009}) who argue that there is no true quasi-periodicity, just a variability process with a featureless red noise power spectrum (e.g. a power spectrum $P(f) \sim f^{-2}$ where $f$ is frequency). A stochastic source with a red noise power spectrum has higher variability at longer timescales and could potentially be responsible for the structures on longer timescales seen in the real lightcurves. The authors suggest that apparent flare peaks may simply be the highest observed flux excursions in such a purely stochastic source and are not isolated events. 

A main reason for the two rather contrasting interpretations of the variable emission from Sgr~A* has been that the nature of the faint emission from Sgr~A* and its relationship to the high flux emission is uncertain. The NIR emission from the Galactic Center is dominated by the central cluster of bright stars, and adaptive optics at 8-meter class telescopes is required in order to separate the faint source Sgr~A* from the closest S-stars (even at this resolution Sgr~A* is still on occasion confused with a relatively bright star). Additional, faint stars may be present very close to Sgr~A* which have not yet been tracked and identified as stars from astrometric monitoring programs \citep[e.g.][]{Gillessen2009}.  While the dramatic high flux variability can be unambiguously attributed to the black hole, when a faint source is detected at the position of Sgr~A*, it is not necessarily clear that the source is Sgr~A*, faint stars, or a combination of both. 
Accordingly, it is not clear whether Sgr~A* continues to emit at all at low fluxes, whether it exhibits a `quiescent state' (a non- or weakly active low state), or whether the low flux emission continues to vary constantly with the same statistical properties as the high flux emission. 

\begin{deluxetable*}{cccccc}
\tabletypesize{\scriptsize}
\tablecolumns{6}
\tablewidth{0pc}
\tablecaption{Flux comparison of bright/notable/faint Ks-band fluxes from Sgr~A* reported in the literature}
\tablehead{\colhead{} & \multicolumn{5}{l}{  } \\
Reference & Obs. & Reported flux  & Photometric cal. & Extinction & Rescaled flux \\
 & Date & [dered~mJy] & rescaling factor  & rescaling factor & [dered~mJy]\\
}
\startdata
 &  &  &  &  & \\
\emph{Brightest States} &  & \emph{Peak fluxes} &  &   &  \\
  &  &  &   &  & \\
Gen+03 & 15 Jun 2003 & 13.2 (10.5+2.7) & 1.10 & 0.76 & 11.0 \\
 &  &  &  &  &   \\
Gen+03 & 16 Jun 2003 & 10 (7.3+2.7) & 1.10 & 0.76 & 8.4 \\
 &  &  &  &  &  \\
Mey+07 & 6 Oct 2003 & 22$^{a}$ & 1.2 & 0.76 & 20.1 \\
 &  &  &  &  &  \\
Tri+07/Mey+06 & 31 May 2006 & 16/23 (+S17) & - & - & (16.7) 13.5$^b$\\
 &  &  &  &  &  \\
Hor+07 & 31 Jul 2005 & 11.6 & $\sim$1.06$^{c}$ & 0.52 & 6.4 \\
 & 2 May 2006 & 26.8 & '' & '' & 14.8 \\
 & 17 Jul 2006& 6.8 & '' & '' &  3.7 \\
 &  &  &  &  &  \\
Do+08 & 3 May 2006 & 0.8 & $\sim$1.06$^{c}$ & - & 8.5 \\
 & 20 Jun 2006 & 0.65 & '' & '' & 6.9 \\
 & 21 Jun 2006 & 0.4 & '' & '' &  4.2 \\
 & 17 Jul 2006 & 0.3 & '' & '' &  3.2 \\
 & 18 May 2007 & 0.6 & '' & '' &  6.4 \\
 & 12 Aug 2007 & 0.2 & '' & '' &  2.1 \\
  &  &  &  &  &  \\
Eck+08 & 15 May 2007 & 24 (+S17) & 0.76 & 0.76  & (13.9) 10.7 \\
 &  &  &  &  &  \\
this work & 5 Aug 2008 & 30.7 (+S17) & - & - &  (30.7) 27.5 \\
 &  &  &  &  &  \\
\hline
 &  &  &  &  &  \\
\emph{Faintest States} &  &    &  &   & \\
 &  &  &  &  &  \\
Hor+02 & 9 May 2001 & $0.09 \pm 0.005$~mJy & $\sim$1.06$^{c}$ & - & $0.95 \pm 0.05$ \\
 &  & (upper limit, not dered) &  &  & \\
 &  &  &  &  &  \\
Sch+02 & 5 Jun 2006 & $2\pm1$~mJy & ? & 0.76 & $1.5 \pm 0.8$ \\
 &  &  &  &  &  \\
Do+08 & 2006: May 3, & $0.192$~mJy &  \\
 & Jun 20, 21, Jul 17 \& & (median flux, not dered)  & 1.06$^{d}$ & - & 2.0 \\
 & 2007: May 18, Aug 12 & $0.082\pm0.017$ & 1.06 & - & $0.9\pm0.2$ \\
 &  & (faintest flux, not dered) & &  &  \\
 &  &  &  &  &  \\
Sab+10 & 23 Sep 2004 & 2.4~mJy & $0.88^{e}\times 0.90^{f}$ & 0.76 &  1.4\\
 & & (upper limit, & & &  \\
 & & no stellar contam.)& & &  \\
 & 23 Sep 2004 & 0.9~mJy & & & 0.5 \\
 &  & (upper limit,& & &  \\
 &  & full stellar contam.) & & &  \\
\enddata
\tablecomments{Reported fluxes from the literature for high and low flux states of Sgr~A* in the literature, rescaled to match the photometric calibration and extinction used in this paper (where S65 has $m_{Ks}=13.7$ and $A_{Ks}=2.5$; see text for details). For the rescaled fluxes, brackets denote the raw observed flux (including S17) and the value without brackets the S17-subtracted estimate of the intrinsic flux. Note that without more detailed analysis we have only been able to reasonably account for and subtract the contribution of S17, and not any fainter stars: thus in the rescaled fluxes quoted here there may still be order of $1-2$~mJy stellar contribution to the flux.\\
 $^a$ though the lightcurve shows a peak flux more like $\sim$24~mJy, this is likely an overestimate since the peak was only observed in one polarization filter (45$^\circ$; see Figure 2 in \citet{Meyer2007}. We take instead 22~mJy, which appears to be a better estimate of the integrated flux (i.e. $F_{45^\circ}+F_{135^\circ}$), according to the modeling of \citet{Meyer2007}.\\
$^b$ we quote the value from our own photometry; this observation night was part of our 2004-2009 dataset.\\
$^c$ Scaling for -0.06 offset in absolute photometric calibration, assuming same calibration as \citet{Do2009b}.\\
$^d$ Scaling for -0.06 offset in absolute photometric calibration (as determined from comparing magnitudes reported in \citet{Do2009b} and \citet{Gillessen2009}, see also \citet{sab10}.\\
$^e$ Scaling for 0.14~mag offset in absolute photometric calibration (as determined from comparing magnitudes reported in \citealt{sab10} and \citealt{Gillessen2009}).\\
$^f$ Scaling for apparent different zeropoint.\\
 \label{table2}}
\end{deluxetable*}

In addition to this, an unbiased overview of the properties of the near-infrared emission from Sgr~A* can be difficult to obtain from the published literature because of publication bias (bright events have individual interest and are often published alone). However, some studies have looked at the statistical properties. \citet{YusefZadeh2006_InfraredFlares} and \citet{yus09} presented lightcurves and flux distributions for Sgr~A* for about $\sim$11 hours of 1.6$\mu$m and $\sim$32 hours of 1.7$\mu$m data observed with NICMOS on the Hubble Space Telescope (HST). With the resolution of the HST the close stellar sources are not as well separated from Sgr~A* as with the VLT or Keck Telescopes, and the stars S17 and S2 overlap with the Sgr~A* source in these observations. The flux distributions were fitted with a Gaussian at low fluxes, which was attributed to the observational noise on constant sources (the contribution from S2, S17 and possible quiescent emission) and a power-law at high fluxes, which was attributed to transient flares. The best fit models implied that Sgr~A* was active (above the noise at low levels) more than 40\% of the time. 

\citet{Do2009} presented an analysis of six nights of K'-band (and one L'-band) observations at the Keck Observatory, using an unbiased set of observations taken between 2005 and 2007. A source at the position of Sgr~A* was always detected in this dataset, with an estimated maximum 35\% contribution from stellar contamination. These authors reported that the source Sgr~A* was continuously variable, based on the larger variance of Sgr~A* compared to stars of similar brightness on five of the six K'-band observation nights. This was the data set used to investigate timing properties of Sgr~A* in which it was claimed the data set was consistent with a featureless red noise power spectrum with no quasi-periodicity.
However, with a sum duration for the K-band observations of about 12.1 hours, this data set did not sample well the higher fluxes of Sgr~A*, i.e. the source was relatively faint compared to publications where variable emission with suggestive quasi-periodic structure have been reported \citep[][for a comparison of Ks-band peak emission from the literature see Table \ref{table2} in the Appendix]{Genzel2003,Trippe2007,Eckart2008_NIRpol}. 
Although the studies of \citet{YusefZadeh2006_InfraredFlares}, \citet{yus09} and \citet{Do2009} have gone some way towards understanding the statistical properties of Sgr~A* in the near-infrared, there has not yet been a study on a very large, unbiased dataset of the variability of Sgr~A* where the rare high fluxes are also well sampled. 

In this paper, we analyse the Ks-band flux distribution of Sgr~A* for the years 2004-2009 from 117 observation nights carried out with the VLT in Paranal, Chile with the aim of seeking the flux-dependent characteristics of the variability of Sgr~A* at both high and low fluxes.  We do this through investigation of the flux distribution of Sgr~A*. The dataset of this paper is $\sim$12 times larger than the data set of \citet{Do2009}. 
In order that our results might be compared with other publications, we give a summary in Table \ref{table2} (in the Appendix) of the brightest and/or notable Ks/K'-band variable emission reported in the literature, as well as the faintest values or upper limits. All are reported in the literature with different calibrations/extinction values, and we have done our best to determine the corrections to scale them to the calibration and extinction values used in this paper.

In Section \ref{section_dataset} we present our observations and the results of using two different methods of photometry: a six year lightcurve (aperture photometry), and the 2009 data set (PSF photometry). In Section \ref{section_results} we present the flux distribution of Sgr~A* and our results from various model fits to the flux distribution. In Section \ref{section_discussion} we discuss our results in the context of two variability states for Sgr~A*: a low-level lognormally varying \emph{quiescent state} and sporadic high flux \emph{flares}. We summarize in Section \ref{section_conclusions}.

\section{Data} \label{section_dataset}

\begin{figure*}[t] 
\begin{center}
 \includegraphics[width=14cm]{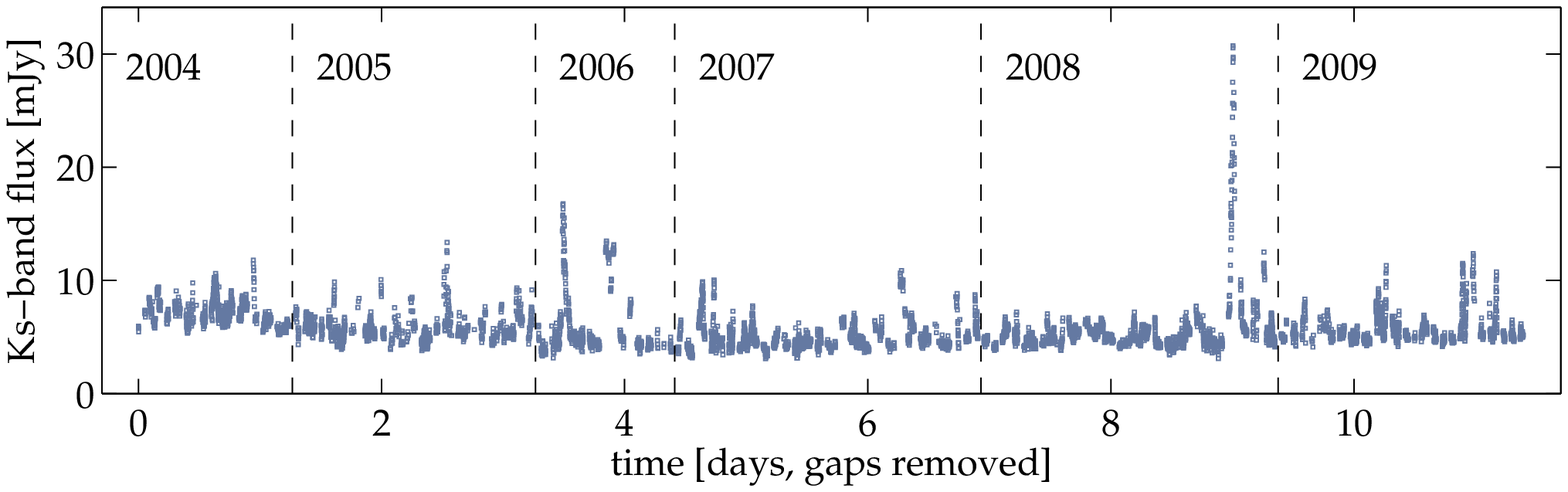}
 \includegraphics[width=14cm]{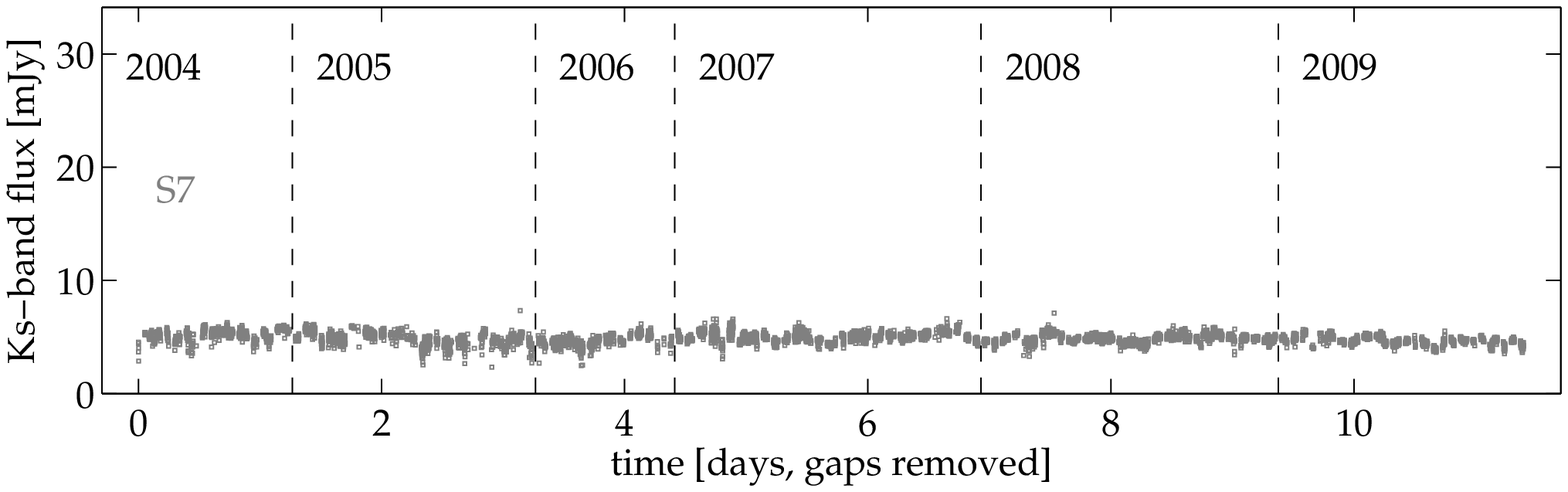}
 \includegraphics[width=14cm]{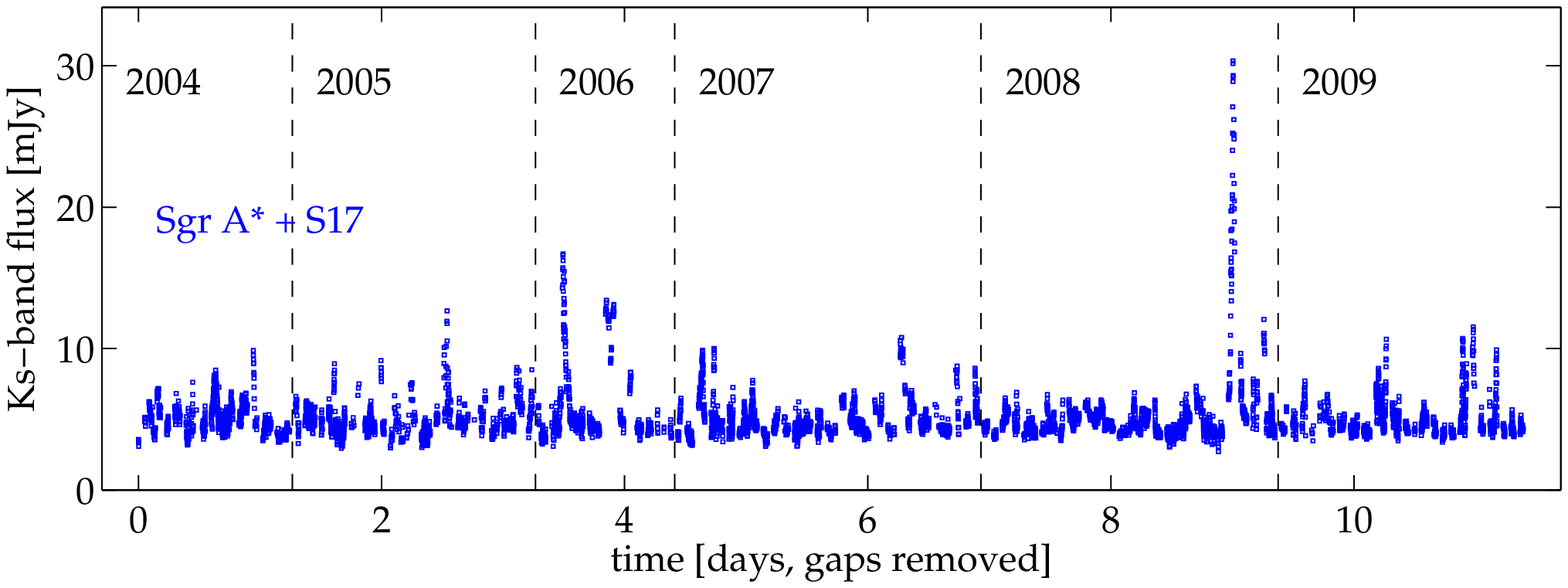}
\end{center}
  \caption{Lightcurve of Sgr~A* 2004 to 2009 from NACO Ks-band observations.  
The top panels shows the lightcurve of Sgr~A* + S17, produced with aperture photometry, versus time with all gaps longer than 0.1 day removed.
The second panel shows the lightcurve produced from the same data and with the same method for S7, a star of similar flux to the Sgr~A*+S17 when faint. The source Sgr~A*+S17 is more variable at low levels than the comparison source S7. In the lowest panel we show the detrended lightcurve, computed from subtraction of difference of the median value from each year, and the year of lowest median value (2006). We do this because there is good justification that the longest timescale trends of the source are dominated by additional faint passing stars (and in the case of 2004, extra flux in the aperture from the halo of S2 which was much closer in that year). Subtracted offsets are roughly 2.2~mJy in 2004, 0.9~mJy in 2005, 0~mJy in 2006, 0~mJy in 2007, 0.3~mJy in 2008 and 0.6~mJy in 2009.\label{fig_dataset}}
\end{figure*}

Since 2002 we have observed the Galactic Center with the near-infrared adaptive optics-assisted diffraction limited imager NACO at the VLT \citep{Lenzen2003,Rousset2003}. Much of the observation time 2002-2009 was spent concentrated on the central few arcesconds, measuring the positions of the S-stars and monitoring Sgr~A*. By now we have amassed a large dataset with which we can investigate statistical properties of Sgr~A*'s variability. While we utilised L' (3.8$\mu$m), Ks (2.18$\mu$m) and H (1.66$\mu$m) bands, by far the largest proportion of our data is taken in Ks-band, collected in either 13~mas pixel scale imaging, 13~mas pixel scale polarimetry, or 27~mas pixel scale imaging mode. 

The presence of many stars close to the central black hole complicates the attempt to acquire accurate photometry for the near-infrared Sgr~A* source. Sgr~A* is usually fainter than the surrounding S-stars, and the stars \emph{move}. A star can on occasion get so close to Sgr~A* in projection that the two sources become confused. For example, the $\sim$16 year orbit star S2 \citep{Ghez2008,Gillessen2009} was confused with Sgr~A* as it passed pericenter in 2002. More recently, the $m_{\rm Ks}\approx15.8$~mag star S17 was confused with Sgr~A* in 2006-2008.

We use two different methods in this paper: 
\begin{enumerate}
\item We first attempted to obtain as large a homogeneous, unbiased, dataset as possible, in order to obtain a best overview of the statistical properties of the variability of Sgr~A*, and to address in particular the variability at (the rarer) high fluxes.  To do this we used aperture photometry because this allowed us to address the most important complication to the photometry of Sgr~A* in 2004 to 2009 -- the star S17, confused with Sgr~A* in 2006-2008. To deal with S17 we used a two-aperture photometry method to determine the combined flux of Sgr~A* and the star S17 in all years 2004 to 2009 with comparable accuracy. This is important because to understand the flux distribution we have to understand the photometric errors at low fluxes. If all years are measured with comparable accuracy the long-term errors can be well approximated with a Gaussian (with width dependent on flux; see Section \ref{Section_obserrors}). We did not include the Ks-band data from 2002-2003 because the star S2 was so close to Sgr~A* during this time that it contributed to the flux measured with this method. Since S2 is much brighter than S17 the flux of Sgr~A*+S17 could not be measured with an accuracy comparable to the other years.
\item
 We secondly looked at a subset of the data in more detail in order to address the nature of the variability at low fluxes. Determining the nature/existence of low level variability (and in particular to distinguish it from observational errors) requires more precise photometry than could be achieved with aperture photometry, in particular with the inclusion of S17. We address the question of the nature of the low level variability by analyzing the (high quality) 2009 data, in which S17 is not confused with Sgr~A*, in greater detail with a PSF-fitting photometric method Starfinder, \citealt{Diolaiti2000}). The flux errors for this method are much smaller and allow us to distinguish with more certainty between true variability and flux errors at low fluxes; this data set is not completely unbiased since Sgr~A* was only reliably detected in 87\% of the selected images. 
\end{enumerate}
By using these two methods in combination, we can overcome the individual difficulties of the methods and piece together a consistent picture of the near-infrared variability of Sgr~A*.

\subsection{Aperture Photometry of Ks-band data 2004-2009}\label{sec_apphot20042009}

Our 2004-2009 Ks-band data set consists of $\approx12 000$ images\footnote{proposal IDs: 072.B-0285, 073.B-0084, 073.B-0775, 073.B-0085, 271.B-5019, 075.B-0093,  077.B-0014, 078.B-0136, 077.B-0552, 078.B-0136, 082.B-0952 and Large Programs 179.B-0261/.B-0932 and 183.B-0100.}. The data were reduced in the standard way by applying a sky subtraction, a flat field correction and a hot/dead pixel correction \citep[see e.g., ][]{Trippe2007}.

We found it necessary to apply a quality cut to eliminate the worst data and ensure we obtained a homogeneous data set on which we could perform accurate photometry. This quality cut was carried out by eye, and the criteria for elimination included: 
\begin{itemize} \item[(i)] the two calibration stars were not in the image (this happened rarely, and only in the polarimetric data) \item[(ii)] image ghosts close to Sgr~A* or PSF artefacts \item[(iii)] simply a bad quality image mostly corresponding to low Strehl ratio data resulting from poor atmospheric conditions. The quality of images for this criterion was judged by the general visibility of S-stars, not by the visibility of Sgr~A* to avoid bias towards bright fluxes from Sgr~A*. 
\end{itemize}
We additionally eliminated 854 images from July in 2009 which were taken in a triggered mode, so that our dataset remains unbiased and representative of the overall variability of Sgr~A*. The remaining data set totaled 6774 images which were used to obtain a lightcurve for Sgr~A*. 

To extract fluxes for Sgr~A* as well as for several control stars, we carried out aperture photometry on each image. However, if we used a standard circular aperture centred on Sgr~A*, we would have a varying contribution from S17 (the brightest star to be confused with Sgr~A* 2004-2009) to the lightcurve as the star moves through the aperture during 2006-2008. A second star within the aperture, but not centered within the aperture also increases the error on our measurements of the flux of Sgr~A*. We would obtain more accurate results if S17 could somehow also be well-centered in the aperture at all times, as well as Sgr~A*. 

To solve the problem of S17 we thus used a two-aperture method with two circular sub-apertures, one centered on Sgr~A* and one centered on S17 and measure their combined flux, averaging the results where the two sub-apertures are each 40, 53 and 66~mas in size. We used this method for all data 2004 to 2009, so that we always measure the combined flux of S17 + Sgr~A*. While there may be additional effects due to confusion with other, fainter stars, at first order our method ensures that we measure the fainter fluxes of Sgr~A* with similar measurement errors from year to year than we would have if we neglected to include a sub-aperture about S17.

To further ensure the self-consistency of our dataset, we used the same set of calibrator stars for calibration of each image. This restricted us to only two suitable calibration stars, S30 and S65, due to the small field of view of some (especially polarimetric) images. To measure the raw counts of control stars and the calibration stars we used the same aperture as for Sgr~A* and S17, but centered the star in just one of the two sub-apertures. For each image, the raw counts of Sgr~A*/S17 and the control stars are then divided by the counts of the two calibrator stars and the calibrated counts are then flux calibrated by scaling relative to the median of S65 for the observation night, using $m_{\rm Ks} = 13.7$~mag for S65 \citep[e.g.][]{Gillessen2009}. Our photometric calibration is also consistent with \citealt{Schoedel2010} (S65: $m_{\rm Ks}=13.64\pm0.02\pm0.06$~mag). To convert magnitudes to fluxes (mJy) we used the zeropoint from \citet{Tok2000} ($m_{\rm Ks}=0$~mag corresponds to 667 Jy). 

 We computed the expected positions of Sgr~A* and of the control stars using the orbital (polynomial) fits of \citet{Gillessen2009}. For polarimetric data, we added the raw photon counts recorded in the apertures for all targets in both ordinary and extraordinary images first in order to recover the unpolarized intensities\footnote{When using NACO in polarimetric mode, images in two polarization filters are recorded at once on the detector by the use of a Wollaston prism plus mask: an ordinary image (in a chosen polarization filter) and an extraordinary image (at a polarization of 90$^\circ$ from the ordinary image).}, before the flux calibration was carried out. We dereddened the final flux values using $A_{\rm Ks} = 2.5$ \citep[][Fritz 2010 in prep.]{Schoedel2010}. Note that the absolute photometric calibration and extinction correction is not important for the analysis of relative fluxes presented in this paper, where we have focused on maximising the long-term relative accuracy of our photometric measurements. However, where we compare with previous publications, we additionally scale the reported fluxes to the photometric calibration (as best as possible) and extinction of this paper.
 
The lightcurve we produced in this way for Sgr~A* is shown in Figure \ref{fig_dataset}. The data presented here is representative of the overall variability of Sgr~A* since it is not biased towards the presence of obvious `flare events' and the only data selection carried out concerned data quality and photometric consistency as explained above. It is by far the most extensive dataset for Sgr~A* that has been published: the observations we present here consist effectively a $\sim$184 hour long lightcurve (if one subtracts from the total time all gaps larger than about 20~min, i.e. gaps not due to sky observations). The total amount of exposure time (i.e. the time spent with the shutter open, excluding overheads) is smaller, and totals $\approx$72 hours, though 184 hours is a better indication of the amount of continuous coverage.

\begin{figure}[t] 
\begin{center}
 \includegraphics[width=8cm]{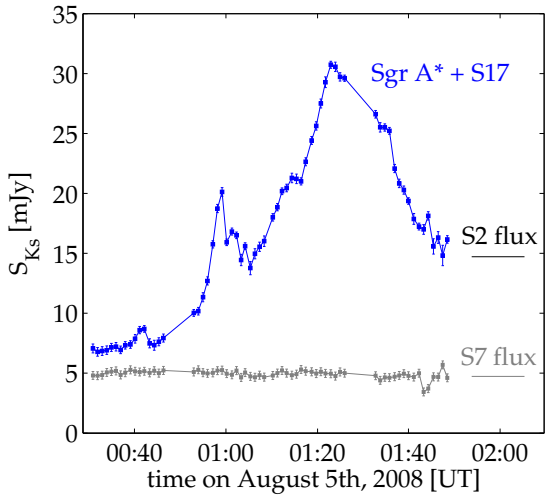}
\end{center}
 \caption{The August~5th, 2008, lightcurve. Sgr~A* + S17 is shown in blue, and the lightcurve of the star S7 is shown for comparison. The flux has been dereddened with $A_{\rm Ks}=2.5$. This flare is the brightest Ks-band flare that has ever been observed; the source reached an intrinsic peak flux of 27.5~mJy, a factor $\approx$1.7 the flux of the star S2. While the high flux excursion is preceded by what looks like a flatter, 'background' level of emission, this is however, at$\sim$7~mJy, several mJy brighter than the low flux levels from most other observation nights,
 and indicates there was increased source activity even at this time. The long-term trend for 2009 has not been subtracted from the lightcurve shown in this plot, and the last six datapoints did not make the data quality cut for the 2004-2009 flux distribution. }\label{fig_superflare}\end{figure}

Figure \ref{fig_dataset} shows the 2004-2009 lightcurve of Sgr~A* and comparison star S7. On top of the more rapid variability, there is a general trend for the lowest fluxes of the lightcurve to wander from year to year. It is at its lowest in 2006. In 2004 the extra flux appears to be due to a combination of a faint star\footnote{identified as S62 in \citet{sab10}, though we disagree with this identification (S. Gillessen, private communication)}, a contribution from the halo of S2, which was much closer to Sgr~A* during that year, and S19, which was confused with S17. The lightcurve also shows some increased flux in 2008 and 2009; and as we will show in Section \ref{sec_stellarcontamination}, there is at least one additional (previously unidentified) faint star which must also contribute to the increased flux in those years.

Since we can generally identify that stars are responsible there is good justification for subtracting these long (i.e. year) timescale trends from the lightcurve. For want of a better method of determining the stellar contribution to the lightcurves we subtracted the difference between the median of the data from each year and the minimum median value (in 2006). This brings the offset in the lightcurve to roughly 3.5-4~mJy, much of which can be reasonably attributed to S17 (between 2.9 and 3.3~mJy in the lightcurve can be attributed to S17; see Section \ref{Section_PSFphot}). 

It is apparent from Figure \ref{fig_dataset} that Sgr~A* is much more variable than the comparison source S7.
Some of the flux excursions are much more dramatic than others, with fluxes above 8~mJy ($\sim$5~mJy intrinsic upon subtraction of S17). Many of these have been previously published  -- for example, the second brightest peak of the dataset, in May~2006, was published previously in \citet{Meyer2006} and \citet{Trippe2007} and has 16.7~mJy ($\sim$13.5~mJy intrinsic; see Table \ref{table2} in the Appendix). 

On 5th August 2008, we saw a particularly extreme event which can be seen as the most dramatic flux excursion in the lightcurve in Figure \ref{fig_dataset}. This event is also shown in higher time resolution in Figure \ref{fig_superflare}. On this particular night the source at the position of Sgr~A* brightened by a factor $>4$ in a period of $\sim$40~minutes, reaching a peak flux of 30.7~mJy (intrinsic flux of $\approx 27.5$~mJy minus S17, a factor increase over the upper limits on the lowest fluxes from Sgr~A* of $\approx$27).
An increase in Ks-band emission of this brightness is unparalleled in the literature for Sgr~A*: the second brightest that has been published is a flare from October 2003 \citep{Meyer2007} which reached an intrinsic flux of $\sim$20.1~mJy (see Table \ref{table2} in the Appendix). 

\subsubsection{Faint Stellar Confusion 2008-2009} \label{sec_stellarcontamination}

\begin{figure}[t]
\begin{centering}
\includegraphics[width=8cm, angle=0]{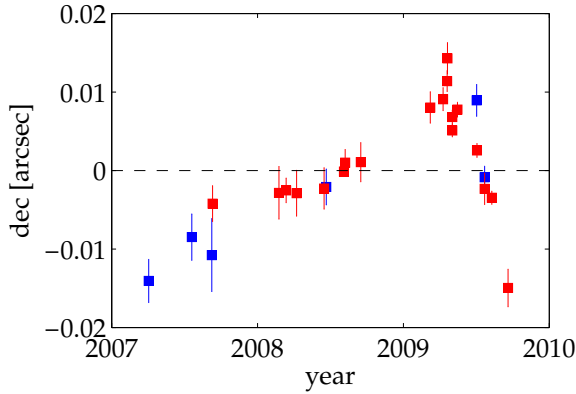}\\
 \caption{Measurements of the source position in declination for Sgr~A* measured from (red) Ks-band and (blue) H-band imaging mosaics between 2007 and 2009. The clear deviations show that there was at least one contaminating star that may have been undergoing a close pericenter passage. Note that the measured deviations are dependent on the intrinsic fluxes of Sgr~A*, which is variable, and the contaminating star. The zero position of the coordinate system is uncertain on the order of $\sim$2-3~mas; for the purposes of this plot the zero position of declination was fixed to the position for which the brightest flux from Sgr~A* was recorded (the $27.5$mJy, $m_{\rm Ks}=13.5$~mag, event from August 5th, 2008). \label{fig_contaminatingstar}}
\end{centering}
\end{figure}

To create the 2004-2009 flux distribution of Sgr~A* we subtracted a trend from the data that we suspected to be due to faint stars. However the trend for higher fluxes in 2008-2009 is not explained by the presence of any known (previously identified) stars. 

Here, however, we are able to present evidence that there was indeed one to two faint stars confused with Sgr~A* in the years 2007 to 2009. In Figure \ref{fig_contaminatingstar} we show that there are deviations in the source position of Sgr~A* measured from imaging mosaics taken between 2007 and 2009. The systematic deviations imply there was at least one, previously unidentified, faint star confused with Sgr~A*. In fact in new, very high quality observations of early 2010, it is possible to identify two new faint stars very close to Sgr~A*. One or both of these stars may have undergone a close pericenter passage. 

From Figure \ref{fig_contaminatingstar} it appears that one of the stars was present in 2007 also, while our offset for 2007 does not argue for any extra star. However as we discuss in Section \ref{Section_previousmeasurements} there appears to be a relatively steady (on the order of $\sim$0.5~mJy) stellar contribution to the flux of Sgr~A* (including the years 2006-2007), which, together with the subtracted offsets, is probably the result of exactly such faint stars moving slowly in and out of confusion with Sgr~A*: probably one of the newly discovered stars does indeed already contribute to the flux in 2007 (with the additional increase in flux in 2009 due to the second star).

It is somewhat a mystery why these stars have not been identified previously. The stars responsible for the observed deviations in the source position have not yet been identified in data prior to 2007 when they should have been resolved separately. Perhaps they have until now always been confused with other stellar sources. The star close to Sgr~A* in 2004 \citep[identified as S62 in ][though we disagree with this identification]{sab10} may be a candidate.

\subsection{PSF photometry of Ks-band data from 2009 using Starfinder}\label{Section_PSFphot}

\begin{figure*}[t]
\begin{centering}
\includegraphics[width=16cm, angle=0]{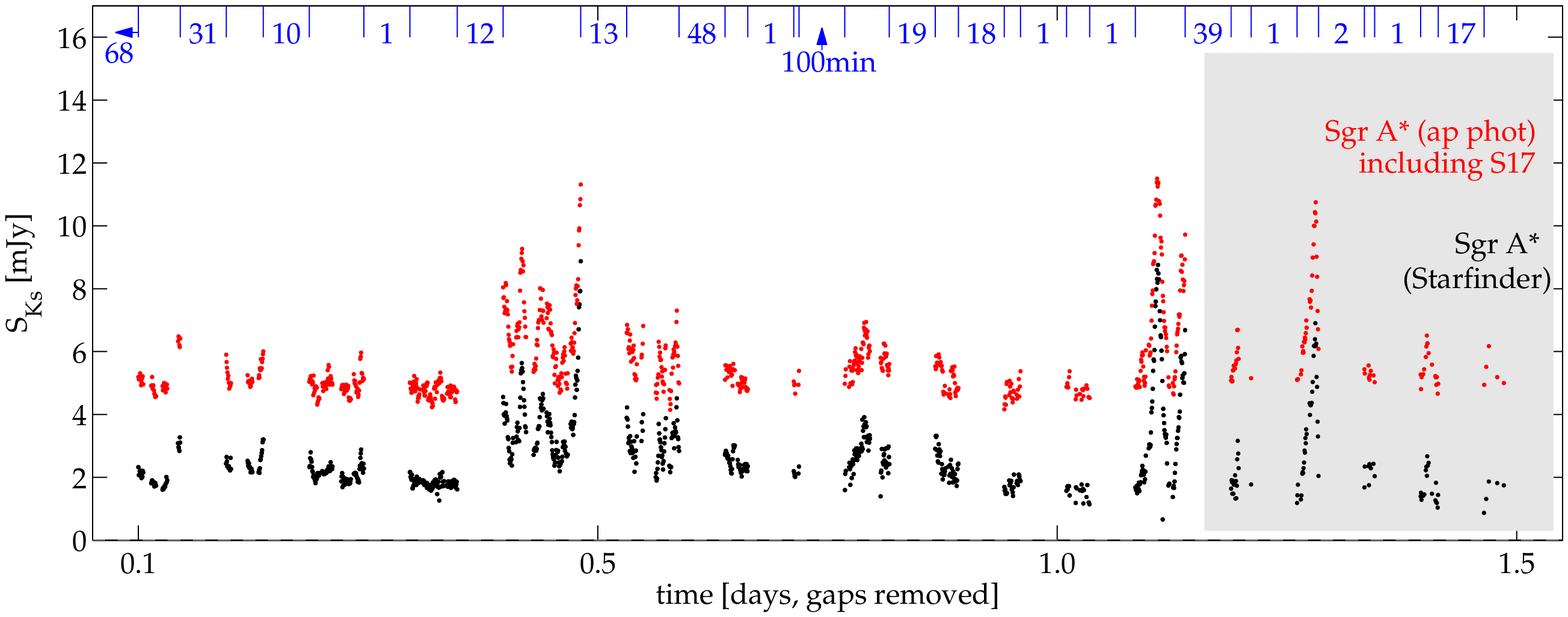}\\
\vspace{-0.4cm}
\includegraphics[width=16cm, angle=0]{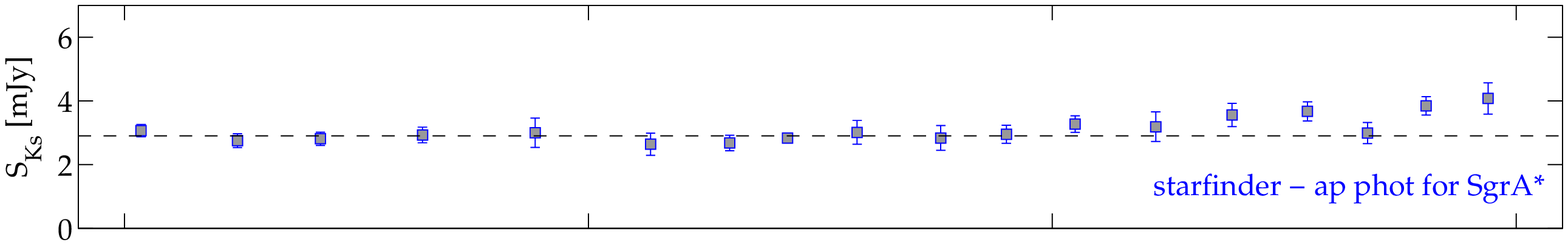}\\
\includegraphics[width=16cm, angle=0]{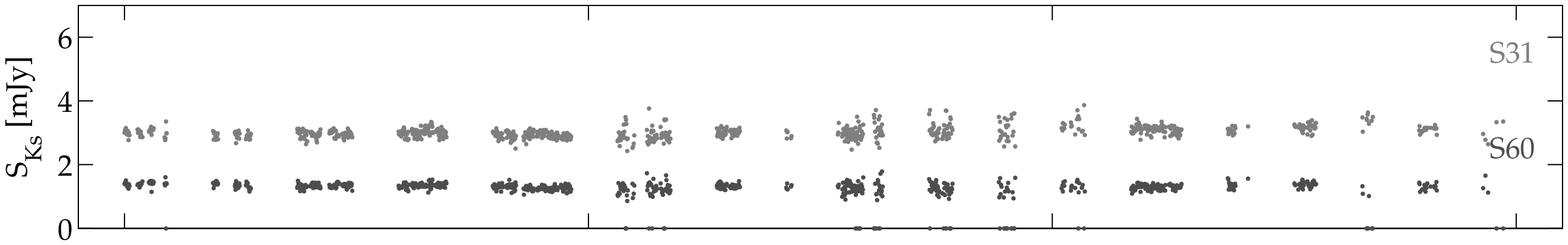}\\
\includegraphics[width=11cm, angle=0]{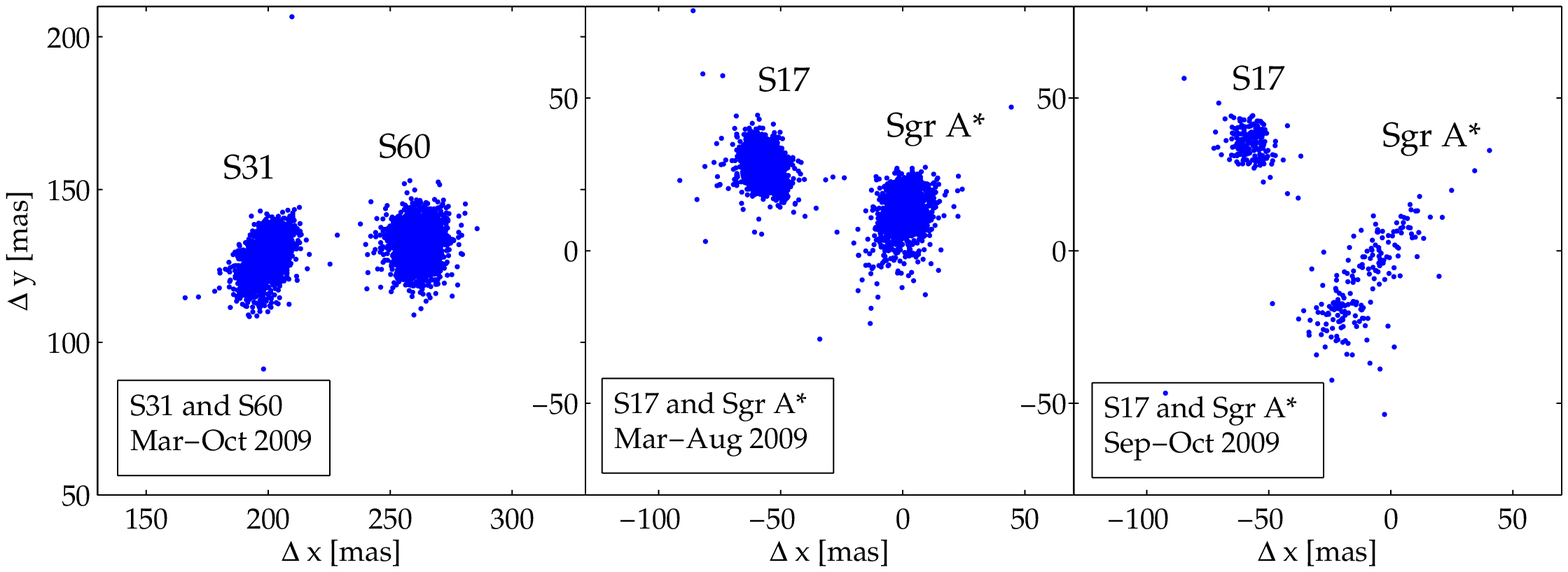}\\
 \caption{\emph{Top:} Lightcurves from aperture photometry (described Section \ref{sec_apphot20042009}) and from Starfinder for 2009 Ks-band, 13~mas pixel scale imaging data. We have not removed here the long-term trend of 0.6~mJy that we did for the 2004-2009 lightcurve. Large gaps in time are removed so that the data can be seen in better time resolution; the real gaps in time between the individual datasets are shown at the top of the figure in days (the first dataset is 68 days from Jan 1, 2009). The data were selected on the basis of whether the star S21 (1.33~mJy) was detected or not, and then only those images (87\%) in which a source was detected within 20~mas of the position of Sgr~A*. Data from September-October 2009, where the faint star confused with Sgr~A* begins to be separated (see bottom panel) is shaded in gray: this data was not used for creating the flux distribution in Section \ref{sec_2009}. 
 \emph{Second from top: Median flux differences between aperture photometry and Starfinder measurements. There is a systematic increase in the offset between the measured fluxes from the aperture photometry method and the Starfinder method towards the end of the year.} 
 \emph{Third from top:} Lightcurves of comparison stars:
 S31 ($2.97\pm0.16$~mJy) and S60 ($1.30\pm0.11$~mJy). S31 and S60 are of similar separation and flux ratio to S17 and Sgr~A* (when Sgr~A* is faint). 
\emph{Bottom:} Detected positions using the Starfinder algorithm on 2009 Ks-band, 13~mas pixel scale imaging data.
S31 and S60 are two stars of similar separation and flux to S17 and Sgr~A* when it is faint. Between March and August both S17 and Sgr~A* are well-detected on the images; from September onwards however the contaminating star (see Section \ref{sec_stellarcontamination}) begins to separate from Sgr~A* and the photometry is unreliable.\label{fig_2009}}
\end{centering}
\end{figure*}

 We additionally carried out PSF photometry on the Ks-band data from 2009 (13~mas pixel scale imaging only) using Starfinder \citep{Diolaiti2000}, which we use especially for investigating the question of whether Sgr~A* is continuously variable at low fluxes (Section \ref{sec_2009}). We use an automated program for running Starfinder on many images developed by \citet{rank2007}. 
The measured fluxes were again calibrated on the S-star S65 ($m_{\rm Ks}=13.7$; \citealt{Gillessen2009}) and the results are shown in Figure \ref{fig_2009}. We note that due to the presence of the two previously unidentified faint stars (Section \ref{sec_stellarcontamination}), we improve our chances of detecting a source in most images, while at the same time the faintness of this additional stellar contribution ensures our flux measurements have very small error, such that this data set is optimum for investigating the variability of Sgr~A* at low fluxes.

To produce the lightcurve we first required that a faint S-star (S21, of $\sim$1.3~mJy) was detected in a given image as a basic data quality cut. From these images we then examined the detected positions of sources in the near neighbourhood ($\sim$100~mas) of the nominal position of Sgr~A* (see Figure \ref{fig_2009}) and selected sources that were within 20~mas of the mean position of this detected source (by this method a source was detected in 87\% of the good quality selected images). Due to this method of source selection, the dataset we present here is not completely unbiased, due to the 13\% of images in which no source was detected, but the fraction in which Sgr~A* was not detected is small enough that the data set still serves well to assess whether Sgr~A* is in general more variable than stars of comparable flux.

From the figure, Sgr~A* certainly appears most of the time more variable than the comparison stars of similar flux (S21 and S60 in Figure \ref{fig_2009}). S60, in particular provides a very good comparison since it is also close to S31, a star of similar brightness to S17; the two sources S60 and S31 also have a very similar separation to Sgr~A* and S17 during 2009 (see Figure \ref{fig_2009}). The flux distribution of S60 is fit by a Gaussian with $\mu=1.309$ and $\sigma=0.098$ which does not give any indication that the photometric accuracy is signficantly worse for sources of this separation and flux ratio (the more isolated star S21 is fit by a Gaussian with $\mu=1.333$ and $\sigma=0.090$). 

In general the Starfinder and aperture photometry fluxes agree very well, with a constant offset of $\sim2.9$~mJy. However, there is a systematic increase in the offset between the measured fluxes from the aperture photometry method and the Starfinder method towards the end of 2009, when the source(s) confused with Sgr~A* has(have) moved the furthest to the south (see Section \ref{sec_stellarcontamination}). This is likely explained by Starfinder's one-PSF fit to the distorted PSF of two sources (which has the effect of decreasing the flux measured with the Starfinder method); the sources are even beginning to become separated at the end of the year (Figure \ref{fig_2009}). 

This method also provides us with an estimate for the flux contribution of S17 to the 2004-2009 flux distribution. The offset of $\sim2.9$~mJy is slightly fainter than the flux of S17 ($3.3\pm0.2$~mJy when separated from Sgr~A* in the 2009 data using Starfinder). The lower total fluxes obtained with the aperture photometry method is probably a result of the higher stellar background surrounding Sgr~A* (i.e. in the background aperture) compared to the more isolated flux calibration star and/or a side effect of the double aperture method. 2.9~mJy is then a low estimate for the contribution of S17 to the combined Sgr~A*+S17 2004-2009 lightcurve, while 3.3~mJy from Starfinder photometry is a high estimate. Therefore we estimate the contribution of the flux of S17 to our 2004-2009 lightcurve produced via aperture photometry to be in the range $2.9-3.3$~mJy.

\section{Results} \label{section_results}

\subsection{Flux Distribution} \label{sec_fluxdist}

For the flux distribution, we define the detection frequency in bin $k$ as
\begin{equation}\text{freq}_{{\rm det},k} = \frac{\Sigma_{i=1}^{N} \Delta t_i(F_k<F_i<F_{k+1}) }{(F_{k+1}-F_k)\Sigma_{i=1}^{N} \Delta t_{i}}.\label{equation_detfreq}
\end{equation}
where $\Delta t_{i}$ and $F_i$ are the exposure time and flux of the $i$th image, respectively, there are $N$ images and $F_k$ and $F_{k+1}$ denote the bin edges. The detection frequency in each bin is divided by the bin width so that the area under the measured flux distribution is equal to 1. In the flux distribution each lightcurve data point was weighted by its exposure time so that the flux distribution is not biased towards observations with shorter time samplings. We chose a logarithmic binning with bins spaced at intervals separated by a factor of $10^{0.05}$. The errors for each bin are computed as the square root of the exposure times in the bin added in quadrature (with the same normalization as in Equation \ref{equation_detfreq}) i.e. \begin{equation} \sigma(\text{freq}_{{\rm det},k}) = \frac{\sqrt{\Sigma_{i=1}^{N} (\Delta t_i(F_k<F_i<F_{k+1}))^2} }{(F_{k+1}-F_k)\Sigma_{i=1}^{N} \Delta t_{i}}.\end{equation} 
The flux distribution of Sgr~A* for the years 2004-2009 as derived from aperture photometry is shown in Figure \ref{fig_fluxdist} with various model fits (described in the next Section), as well as for a comparison star.
The parameters and fit results for the model fits are given in Table \ref{table1}.

\subsubsection{Accounting for observational errors}\label{Section_obserrors}

\begin{figure}[t]
\begin{centering}
\includegraphics[width=8cm, angle=0]{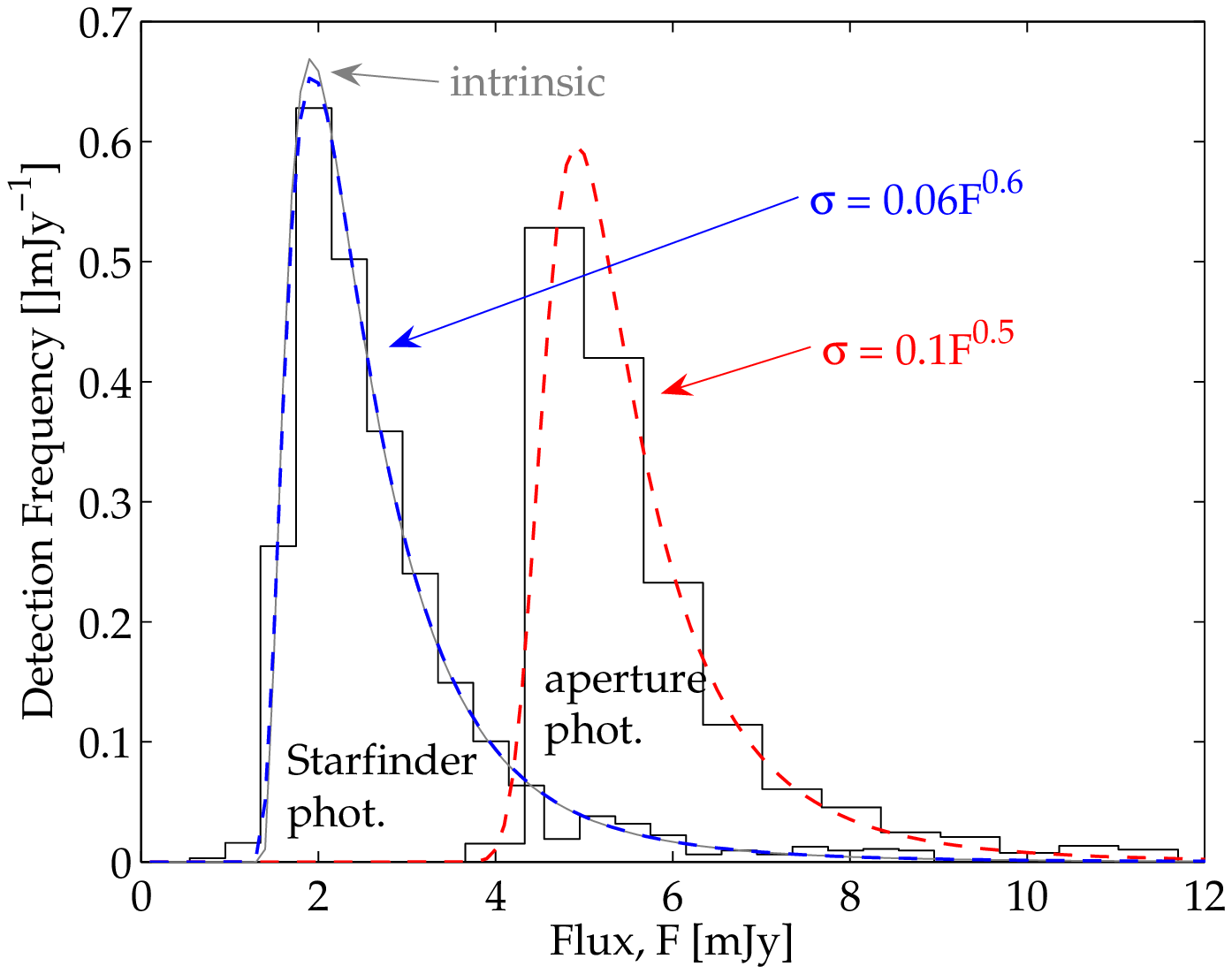}
\includegraphics[width=8cm, angle=0]{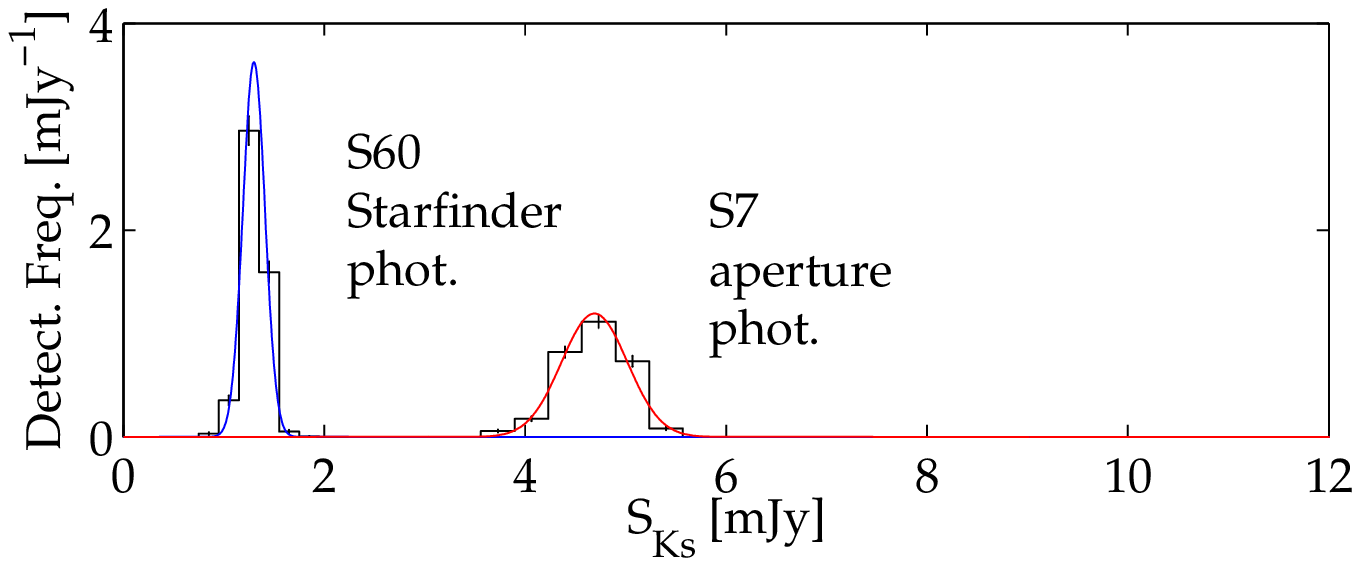}
 \caption{\emph{Upper panel:} Demonstration of the effect of observational errors on the flux distribution. We show the flux distribution from the Starfinder 2009 photometric dataset compared to the flux distribution determined via aperture photometry for the same (2009 subset) dataset. The fluxes obtained from aperture photometry method include the additional flux of S17. The size of the bins is scaled proportional to the typical error at the peak of each distribution. The blue dashed line shows the best fit lognormal model to the Starfinder photometry distribution, convolved with the error law for 2009 Starfinder photometry data $\sigma=0.06F^{0.6}$ (Equation \ref{Equation_SFobserror2009}), which are almost negligible (compare to the intrinsic flux distribution shown in gray). For the same data, but analyzed with aperture photometry, the observational errors are important and have the effect of slightly widening and flattening the peak of the distribution. The red dashed line shown is not a fit to the flux distribution produced from the aperture photometry data, but rather, it is the best fit model for the Starfinder photometry with a constant flux offset and convolved with the best fit error law \emph{determined from stars} for 2009 aperture photometry data (see footnote \ref{footnote_APobserror2009}). \emph{Lower panel}: Comparison stars with fluxes close to the mode the flux distributions shown above, demonstrating the much greater variance in the flux distribution for Sgr A* than that expected from observational noise in the photometric data set obtained with Starfinder (see Section \ref{sec_2009}). 
  \label{fig_comparefluxdists}}
\end{centering}
\end{figure}

The main difficulty with fitting the distribution of fluxes is the influence of observational errors, which are important enough at low fluxes to widen the peak of the distribution from its intrinsic shape.

We characterized the observational errors by fitting Gaussian functions to the flux distributions of 22 nearby stars of fluxes between $\approx$0.6 and 25~mJy. Fitting a power law we obtain
\begin{equation}
 \log_{10}\sigma_{\rm obs}(F) = (-0.76\pm0.08)+ {(0.5\pm0.1)}\log_{10}F.\label{Equation_APobserror}
\end{equation}
similar to that of \citet{Do2009} who found a typical dependence of $\sigma_{\rm obs}(F) \approx 0.2F^{0.3} \text{~mJy}$ for their data. The $\approx$0.5 power-law dependence of the errors is consistent with a photon noise origin \citep{Fritz2010}.
We do not find any flattening of the power law dependence of the noise towards low fluxes with statistical significance.

The 13 mas pixel data from 2009 is of much higher quality and for the Starfinder photometry has noise properties
\begin{equation}
 \log_{10}\sigma_{\rm obs}(F) = (-1.2\pm0.1)+ {(0.6\pm0.2)}\log_{10}F. \label{Equation_SFobserror2009}
\end{equation}
The noise dependence was determined in the same way as for the 2004-2009 dataset through comparison of 22 stars of fluxes 0.5 to 25~mJy.\footnote{the increase in accuracy appears to be due to both the higher quality of 2009 data and the higher accuracy of the Starfinder method: for example, for the same data of 2009 used for the Starfinder data set, but analyzed with the aperture photometry
 method, the noise properties are $$\log_{10}\sigma(F)~=~(-1.0\pm0.1)+(0.5\pm0.2)\log_{10}F,$$ which is a significant improvement on the overall 2004-2009 aperture photometry accuracy, but still not as good as the accuracy obtained with Starfinder for the same dataset.
\label{footnote_APobserror2009}}

\begin{figure*}[t]
\begin{centering}
\includegraphics[width=7cm]{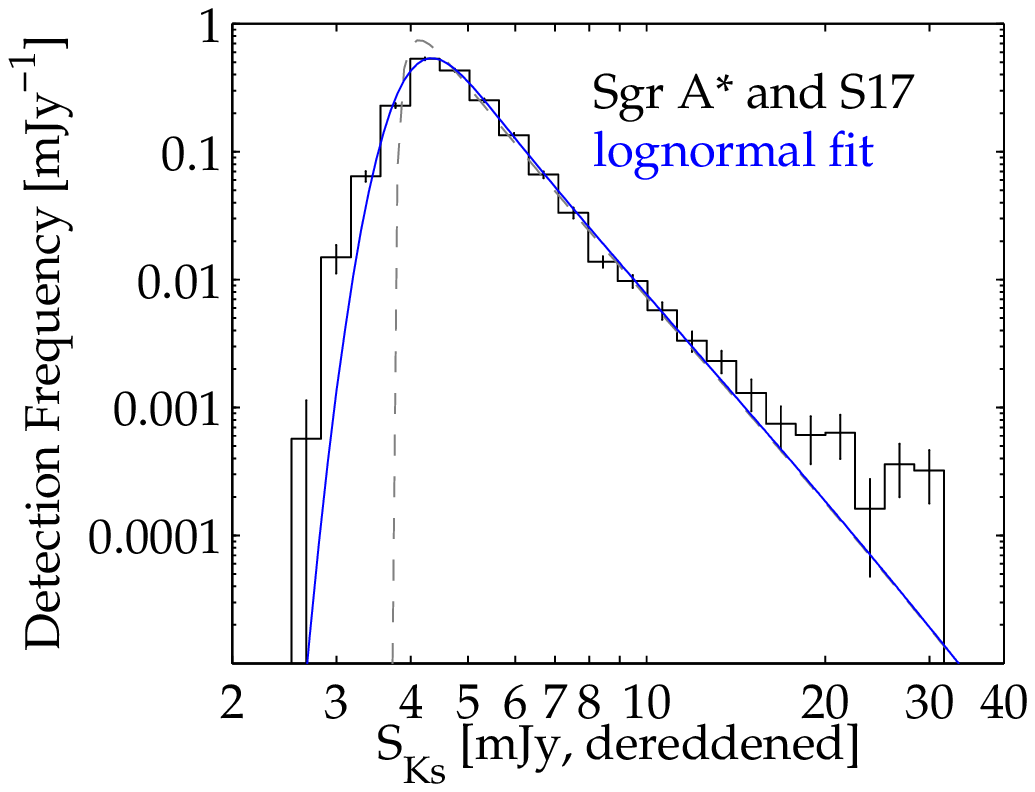}
\includegraphics[width=7cm]{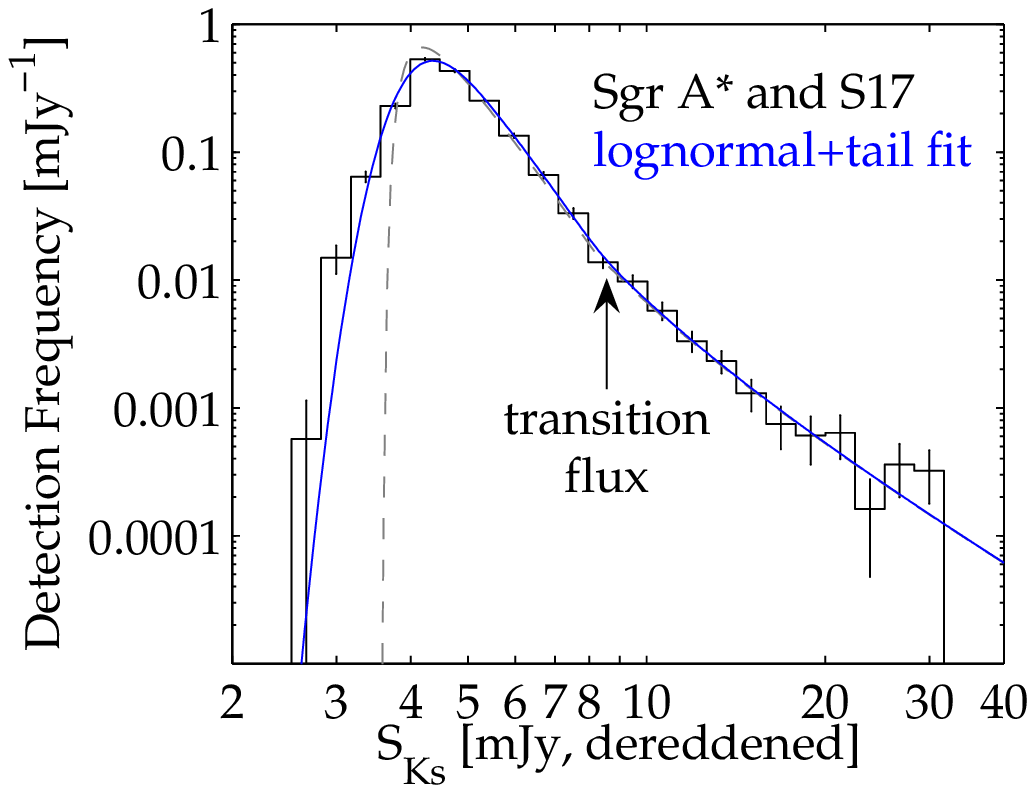}\\
\includegraphics[width=5cm]{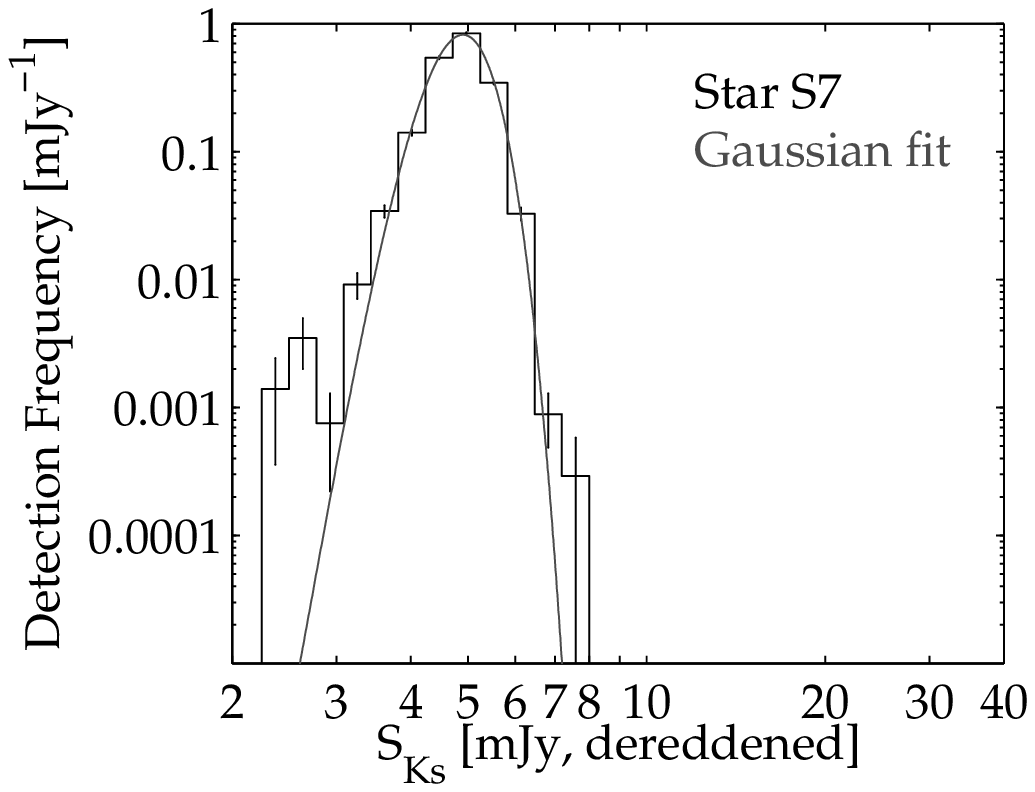}\\
 \caption{{\emph{Left and center:}} Lognormal and lognormal+tail model fits to the 2004-2009 flux distribution (see Sections \ref{Section_logn} and \ref{Section_logntail}). The lognormal+tail model with a transition to a flatter tail towards high fluxes is a significant improvement to the fit, at $>6\sigma$ confidence. Note that although the transition flux appears to be at about 8~mJy (since the contribution of S17 is always added to Sgr~A* in this photometric dataset), the intrinsic transition flux is actually $F_t=4.6\pm0.5$ (see Table \ref{table1}). The blue lines show the full model flux distribution including observational errors ($\sigma(F) = 0.17F^{0.5}$; Equation \ref{Equation_APobserror}), while the dashed gray line shows the intrinsic flux distribution (see Section \ref{Section_obserrors}). Essentially all measurements below $\sim$3.5~mJy occur due to observational errors. The best fit parameters for these models can be found in Table \ref{table1}. \emph{Right:} the flux distribution of comparison star S7, together with the best fit Gaussian. The Gaussian has $\mu=4.9$ and $\sigma=0.48$ for a $\chi^2/{\rm dof}=37.2/8$. The largest contribution to $\chi^2$ (on the order of 20 to 30) for both S7 and Sgr~A* best fit models comes from the low flux wing of the distribution, where a Gaussian fails to be a good model for the observational errors. \label{fig_fluxdist}}
\end{centering}
\end{figure*}

We include the effect of observational errors in our model fits by 
convolving the intrinsic flux distribution with a Gaussian of width $\sigma_{\rm obs}(F)$, which has the dependence on flux of Equation \ref{Equation_APobserror}:
\begin{equation}
P_{0+{\rm err}}(F) = \int P_{0}(F')\frac{1}{\sqrt{2\pi} \sigma_{\rm obs}(F)}\exp\left(\frac{-(F-F')^2}{2{\sigma_{\rm obs}(F)}^2}\right)dF'. \label{Equation_P_lognerrors}\\
\end{equation} where $P_0$ is the intrinsic flux distribution (without observational errors). The effect is only important at low fluxes $\lesssim 6$~mJy (see Figure \ref{fig_fluxdist}) and is responsible for broadening the low flux peak in the observed flux distribution for Sgr~A*. 

\begin{deluxetable*}{cccc|cccccc}
\tabletypesize{\scriptsize}
\tablecolumns{14}
\tablewidth{0pc}
\tablecaption{Fits to the observed flux distribution}
\tablehead{
\multicolumn{4}{c|}{ } &  \multicolumn{6}{c}{ } \\
\multicolumn{4}{c|}{ \textbf{logn+err (see Section \ref{Section_logn})}} & \multicolumn{6}{c}{\textbf{logn+tail+err (see Section \ref{Section_logntail})}} \\ 
\multicolumn{4}{c|}{ } &  \multicolumn{6}{c}{ } \\
$\mu_*$ & $\sigma_*$ & $F_b$ & $\chi^2/{\rm dof}$ & $\mu_*$ & $\sigma_*$ & $F_b$ & $F_t$ & $s$ & $\chi^2/{\rm dof}$ \\
ln(mJy) & ln(mJy) & mJy & & ln(mJy) & ln(mJy) & mJy & mJy &  &
}
\startdata
\multicolumn{4}{l|}{\emph{}} & \multicolumn{6}{l}{\emph{}}\\ 
\multicolumn{4}{l}{\emph{Aperture Photometry on 2004-2009 data}} & \multicolumn{6}{|l}{\emph{Aperture Photometry on 2004-2009 data}}\\ 
& & & & & & & & \\
$-0.12\pm0.04$ & $0.92\pm0.03$ & $3.75\pm0.04$ & $79.7/18$ & $0.05\pm0.06$ & $0.75\pm0.05$ & $3.59\pm0.06$ & $4.6\pm0.5$ & $2.70\pm0.14$ & $47.8/16$ \\
& & & & & & & & \\ \hline
\multicolumn{4}{l|}{} & \multicolumn{6}{l}{}\\ 
\multicolumn{4}{l}{\emph{Starfinder Photometry on 2009 subset}} & \multicolumn{6}{|l}{\emph{Starfinder Photometry on 2009 subset}}\\ 
& & & & & & & & \\ 
$0.04\pm0.06$ & $0.77\pm0.05$ & $1.34\pm0.05$ & $19.0/14$ & $0.07\pm0.06$ & $0.73\pm0.06$ & $1.31\pm0.06$ & $4.8\pm0.9$ & $2.1\pm0.6$ & $15.3/12$ \\
& & & & & & & & 
\enddata
\tablecomments{Best fit parameters and formal fit errors for the large (aperture photometry 2004-2009) and high quality (Starfinder 2009) photometric datasets. For description of the models and the parameters see Sections \ref{Section_obserrors}, \ref{Section_logn} and \ref{Section_logntail}.
Note that $F_t$ refers to \emph{intrinsic flux}, while on the observed flux distribution the break \emph{appears} to occur at a flux of $F_b+F_t$. 
\label{table1}}
\end{deluxetable*}

\subsubsection{Lognormal model}\label{Section_logn}

A lognormal fit to the 2004-2009 flux distribution from aperture photometry is shown in Figure \ref{fig_fluxdist}, with the best fit parameters given in Table \ref{table1}. This model has intrinsic probability distribution $P_0 = P_{\rm logn}$ where
\begin{equation}P_{\rm logn}(F) = \frac{1}{\sqrt{2\pi} \sigma_* (F-F_b)} \exp\left( {- \frac{(\ln(F-F_b) - \mu_*)^2}{2{\sigma_*}^2}}\right)\label{Equation_P_logn}\end{equation} 
where $F$ represents the observed flux (measured in mJy here and elsewhere in this paper), and $F>F_b$ with $F_b$ a flux offset due to some constant contribution (e.g. S17 + contaminating stars). The parameters $\mu_*$ and $\sigma_{*}$ of the lognormal distribution have a natural analogy to the normal distribution when exponentiated: the source can be thought of as having a median flux of $\exp({\mu_{*}})$ with a multiplicative standard deviation of $\exp({\sigma_*})$ (i.e. the interval $\exp({\mu_{*}})/\exp({\sigma_*})$ to $\exp({\mu_{*}})\times(\exp{\sigma_*})$ contains $\approx$68\% of the probability). 

\subsubsection{Lognormal model with flatter high flux tail}\label{Section_logntail}

The residuals at low fluxes of the lognormal model are probably just due to the low flux wing of the error distribution which is not perfectly Gaussian (see Star S7 in Figure \ref{fig_fluxdist}). The residuals for the lognormal model at high fluxes, on the other hand,  can not arise from measurement error, and we obtain a much better fit to the flux distribution with a lognormal model which makes a transition to a flatter tail at high fluxes. The intrinsic probability distribution in this case is
\begin{equation}
 P_{\rm logn+tail}(F) = \left\{
   \begin{array}{lr}
    k P_{\rm logn}(F) & : F \leq F_{b}+F_{t}\\
    k P_{\rm logn}(F_t) \left(\frac{F-F_b}{F_t}\right)^{-s} & : F > F_{b}+F_{t}
   \end{array}
\right.\label{Equation_P_logntail}
\end{equation}
where $P_{\rm logn}(F)$ is the lognormal distribution of Equation \ref{Equation_P_logn}, $F_t$ is the flux at which the distribution makes the transition to the flatter tail, $s$ is the power-law slope of the tail, and $k$ is a renormalizing factor\footnote{$k=1/(\frac{1}{2} + \frac{1}{2}{\rm erf}((\ln{F_t}-\mu_*)/\sqrt{2{\sigma_*}^2}) + F_t P_{\rm logn}(F_t)/(s-1))$} such that the total probability is 100\%.

The flattening of the distribution towards high fluxes in the 2004-2009 aperture photometry flux distribution is significant at $>6\sigma$. However, the fits are not formally `good fits'. This is to the most part due to residuals in the lowest flux bins, most likely caused by the deviation of the observational errors from the plain Gaussian description used in our model. If we instead only fit only to fluxes $>3.5$mJy (ignoring the lowest three bins), we obtain a more reasonable $\chi^2/{\rm dof}$ for the lognormal plus tail model of 17.4/13 (while the best fit lognormal model for the same has 37.6/15). In this case the flatter tail of the distribution is still significant at $>4\sigma$.

Interestingly, in the 2009 Starfinder photometric dataset the tail is also present with similar best fit parameters, although in this smaller dataset (which only reaches fluxes of $\sim$10~mJy) the tail has a significance of only 1.4$\sigma$.

\subsection{Is Sgr~A* a continuously variable source?} \label{sec_2009}

What is not clear in the 2004-2009 data set is whether it is appropriate to fit only continuously emitting, continuously variable models to the flux distribution. A plausible alternative is a model in which the source of near-infrared emission emits only sporadically (i.e. exhibits `on' and `off' periods). If this is the case observational errors alone (on the measurement of a constant component to the flux) should be responsible for observed variations in the flux level during `off' periods.  The goal of this section is to determine whether there is too much variation at low levels for the data to be explained by observational errors in `off periods' of a sporadic model.

We use the following prescription for a sporadic model for the flux distribution:
\begin{eqnarray}
\nonumber P_{0+err}(F) &=&  p_{\rm var} \int P_{\rm 0}(F') P_{\rm gauss}(F-F'-F_b)dF'\\
&+& (1-p_{\rm var})P_{\rm gauss}(F-F_b) \label{Equation_P_noncontinuous} \end{eqnarray}
where $p_{\rm var}$ is the fraction of time the source spends in an `on' state.

Using $P_0(F) = P_{\rm logn}(F)$ (Equation \ref{Equation_P_logn}) and fitting to both the 2004-2009 aperture photometry flux distribution and to the 2009 Starfinder distribution with the measured noise laws of Equations \ref{Equation_APobserror} and \ref{Equation_SFobserror2009}, we find in both cases the best fits to the distributions require $p_{\rm var}>92\%$ and $p_{\rm var}>96\%$ (at $3\sigma$) respectively, i.e. the measured noise laws can not account for the observed amount of variability at low fluxes.

To look at the question from a different angle, we investigate what level of observational noise would be required to produce the observed variation at low fluxes. To do this we allow the normalization of the noise law to be a free parameter in the model fits, i.e. \begin{equation}\log_{10}\sigma_{\rm obs}(F) = A + 0.5\log_{10}F.\end{equation} We find $A=-0.66\pm0.02$ and $A=-0.79\pm0.04$ for the 2004-2009 aperture photometric datset and the 2009 Starfinder dataset respectively. We also refit the noise laws to the stars (see Equations \ref{Equation_APobserror} and \ref{Equation_SFobserror2009}) fixing the slope of the error law to 0.5 and obtain $A_{\rm stars}=-0.84\pm0.05$ and $A_{\rm stars}=-1.12\pm0.07$ for the 2004-2009 aperture photometric datset and the 2009 Starfinder dataset respectively. The best fit noise law to the 2004-2009 flux distribution from aperture photometry allows $p_{\rm var}=68\pm7$ but the normalization of the noise law is inconsistent with that of the stars at a level of 3.3$\sigma$. For the 2009 flux distribution from Starfinder photometry the required noise law is even more unlikely, and is ruled out at a level of 4.5$\sigma$. We conclude that the observed behavior of Sgr~A* at low fluxes is best consistent with a continuously emitting, continuously varying source.

\subsection{Variations on timescales of weeks to months} \label{sec_longtermvar}

We also detect longer timescale trends in the low flux level of Sgr~A* in the 2009 data (Figure \ref{fig_2009}), such as an $\sim$0.8~mJy increase in the minimum level from the fourth to the fifth data sections, with a gap in time of only 12 days, and a similar size decrease from the second to the fourth data sections with a gap in time of only 11 days. 

It might be that a passing star (such as the two new stars discovered in 2010, see Section \ref{sec_stellarcontamination}) is responsible for these trends, either passing in and out of confusion very quickly, or even just moving within the Sgr~A* PSF and distorting the flux: the further apart the two sources the lower the fitted flux. However these stars cannot move quickly enough for this: the change in flux is so large that in either case the star must move by at least $\sim$50~mas within a period of 12 days to distort the flux as much as observed, which at the 8kpc distance of the GC requires an average speed of at least 0.1$c$ over a two week timescale. Even assuming the most extreme case -- a star moving only in the plane of the sky and falling towards Sgr~A* in a parabolic orbit at the escape speed -- a star should take at least $\sim$110 days (0.02$c$ in average), to cover such a distance. A larger centroid shift between the datasets would also be expected if it were the case that a star were distorting the source PSF ($\sim$20~mas instead of the observed $\sim5-8$~mas). Furthermore, even if this could explain the trend between the fourth and fifth datasets, the same explanation would have to be invoked to again explain the variation between the second and fourth datasets, also on a timescale of only $\sim$11 days. 

We therefore conclude that the observed trends on timescales of a few weeks in the lightcurve are most likely intrinsic. Though they can not be due to direct flux contamination from passing stars, the trends are possibly still related to the feeding of the accretion flow from stellar winds of the passing stars, which may trigger increased activity as the star passes orbital pericenter (e.g. \citealt{Loeb2004}).

\section{Discussion/Interpretation}\label{section_discussion}

\subsection{Two states of Sgr~A* in the near-infrared}

We have shown in Section \ref{section_results} that the flux distribution of Sgr~A* is best described by a continuously emitting, continuously variable component with a lognormal flux distribution at low fluxes, but additionally that there is a significantly flatter tail above about 5~mJy, implying that \emph{there is something different about the high flux emission from Sgr~A*}. Put together, it seems justified to identify 
\begin{itemize}
 \item emission below 5~mJy as \emph{quiescent state} emission
 \item emission above 5~mJy as \emph{flaring} emission.
\end{itemize}
This is the (phenomenological) definition by which we hereafter refer to either `flares' or `quiescent' emission in Sgr~A*. 
We present how the two states may fit into a possible physical picture below.

\subsubsection{Quiescent State Emission}

That the low fluxes of Sgr~A* are fit by a continuously present, variable component suggests that there is truly a \emph{quiescent state}\footnote{a \emph{quiescent yet variable state}, in a similar sense to the quiescent state of longer wavelengths which nevertheless exhibits variability (factor 20-100\% in radio to submm regime).} in Sgr~A*, i.e. a low-level, lognormally varying component (factor $\sim$2 in flux of the lognormal component corresponds to 1$\sigma$ variability). Our finding of continuously emitting, continuously variable component confirms the findings of \citet{Do2009} who first reported continuous variability of Sgr~A* from the larger variance of the flux of Sgr~A*, as compared to stellar sources, in 5 out of 6 K'-band observation nights.

Lognormal distributions of fluxes are seen in X-ray binaries, where they are interpreted as a sign that multiplicative processes are behind the variability (e.g. \citealt{Uttley2005}). A good example of a multiplicative process is the model of \citet{Lyubarskii1997} where variability is produced through the inwards propagation of accretion rate fluctuations at different radii. These variations, produced in regions of the flow which themselves do not emit near-infrared emission, all combine multiplicatively as they propagate inwards. The end result of combined fluctuations is then imprinted on the variability produced by the (say, near-infrared) emitting region of the flow. A very similar idea may work with magnetic turbulence, which has been shown in GRMHD/MHD simulations to be capable of producing the factor 40-50\% millimeter variability \citep{Dexter2009,Chan2009,Goldston2005}. 

From where exactly the quiescent emission arises is however a difficult question. Nonthermal electrons are required to emit in the near-infrared, and it's not clear how these should be related to the thermal electrons emitting at submm wavelengths, or how they should necessarily be arranged within the accretion inflow/outflow. \citet[][see also \citealt{Ozel2000}]{yuan03} require a hybrid thermal/non-thermal electron distribution to explain the spectrum of Sgr~A* (in particular the excess radio emission at low frequencies). 
The models predict that the same non-thermal electrons (with $\sim$1.5\% of the thermal energy) could produce emission in the near-infrared. Thus it might be interesting to compare the flux distribution of the NIR quiescent state emission ($F\lesssim 5$~mJy) with that of the low-frequency radio emission, as well as correlations in timing properties, such as whether the medium-timescale trends ($\sim$2 weeks) observed in the 2009 data (Figure \ref{fig_2009}) correlate with any similar long-term trends in the radio regime. More detailed comparison of the low-level NIR emission from Sgr~A* with other wavelengths (e.g. submm) could also shed light on how non-thermal electrons are related to the thermal population.

\subsubsection{Flaring Emission}

It seems unlikely that the same variability process is responsible for both high and low flux emission from Sgr~A*. There is no obvious reason for the underlying process, if multiplicative at low fluxes, to deviate at high fluxes in order to explain the flatter tail above $4.6\pm0.5$~mJy. The flatter tail of the flux distribution in Sgr~A* requires at the very least some transition within the source (to a \emph{flaring state}), triggered around an intrinsic flux of 5~mJy, such that either large fluctuations or their emission undergo some kind of runaway amplification.  

On the other hand the physical mechanism behind the flares could be unrelated to (though possibly still triggered by) the mechanism producing the low level variability. The flare tail could arise from additional sporadic \emph{flare events} which dominate the distribution of fluxes above 5~mJy. In this case 5~mJy is not a physically interesting flux outside the fact that it is the flux at which one is equally likely to observe either a flare event or a low level lognormal variation.

Spontaneous magnetic reconnection is a good candidate for a physical process that could give rise to sporadic flares in Sgr~A* \citep[e.g.][]{yuan03,ding10}. The decrease in magnetic field strength that would accompany the conversion of magnetic energy to high energy electron acceleration in a magnetic reconnection event can explain the non-one-to-one rise in the lightcurves of the so far brightest NIR/X-ray flare from Sgr~A* \citep{DoddsEden2010}.

A number of other properties also indicate that there are two types of NIR emission from Sgr~A*, with differences between high and low flux emission, which include:
\begin{itemize}
 \item \emph{Spectral Index}. There are indications that the spectral index at low fluxes is redder than at higher fluxes \citep{Ghez2004,Eisenhauer2005,Gillessen2006,yus09}, although this has been disputed by other authors \cite{Hornstein2007}.
 \item \emph{Polarization}. Changes in polarization in the rising/decaying phases of flares are seen \citep{eck06}, and low level emission appears to be generally more highly polarized ($\sim30-40\%$) than high flux emission ($\sim 10\%$) \citep{Meyer2006,Trippe2007}.
 \item \emph{Association with X-ray flares}. The best fit model (\emph{Lognormal+tail+errors}) to the 2004-2009 flux distribution has 4.5\% of the probability above the best fit transition flux of $F_t=4.6$~mJy, which, interestingly, is close to the X-ray flaring rate of $\approx4-7\%$ (for flares occurring once a day with a duration of 60-100~min; \citealt{Baganoff2003HEAD}). Estimates of the NIR flaring rate including emission below 5~mJy, on the other hand, typically put the NIR flaring rate much higher at $\approx40\%$ \citep{Eckart2006_FlareActivity,YusefZadeh2006_InfraredFlares}. Perhaps the additional low-level component is even responsible for differences in the NIR/X-ray peak flux ratios.
\end{itemize}

There are a number of high flux Ks/K'-band events ($\gtrsim 5$~mJy), not part of the 2004-2009 dataset presented here but published elsewhere in the literature, which are also of $\gtrsim$5~mJy intrinsic flux and can be identified as belonging to the flare tail of the flux distribution. To list some of the brightest (see Table \ref{table2} in the Appendix for details): $\approx$11 and 8.4~mJy \citep[15, 16~June~2003;][]{Genzel2003}; 20.1~mJy \cite[6~October~2003;][]{Meyer2007}; 13.5~mJy (31~May~2006; \citealt{Trippe2007}, \citealt{Meyer2007}, and part of our 2004-2009 dataset); 10.7~mJy \citep[15~May~2007;][]{Eckart2008_NIRpol}; and 14.8~mJy \citep[][]{Hornstein2007}.

\subsection{Consistency with previous measurements of Sgr~A* at low Ks-band fluxes}\label{Section_previousmeasurements}

Our results are reasonably consistent with the median flux of 2.0~mJy over six separate observation nights from 2006-2007 found by \citet{Do2009} (see Table \ref{table2} in the Appendix for the scaling factors applied to scale the reported values to our calibration/extinction). We find a median flux of $\approx$1.6~mJy for our 2004-2009 lightcurve with the long-timescale trend and S17 (assuming 3.1~mJy) removed (note however that this still includes some constant contribution: the median flux of the lognormal component alone is 1.1 mJy). 

A number of previous measurements have put upper limits on any long-term steady quiescent state at near-infrared wavelengths which have been on the $\sim$1-1.5~mJy level \citep[][see Table \ref{table2} in the Appendix]{Hornstein2002,Schoedel2007,sab10}. These measurements are consistent with the picture we have put forward here for Sgr~A* at low levels, of a continuously emitting, continuously varying source. For a lognormally varying quiescent state with the properties of our best fit model to the 2004-2009 flux distribution, an upper limit of 1-1.4~mJy in selected observations is not surprising: the source is expected to be at $\lesssim$1~mJy roughly $50\%$ of the time. 

However, more likely the upper limits of 1-1.4 mJy include a stellar contribution to the flux. \citet{sab10} estimated an upper limit on the \emph{intrinsic} flux from Sgr A* of $F\lesssim$0.5~mJy in the presence of a likely $\sim$0.8-1~mJy stellar contribution. In the context of our best fit lognormal model, this low flux measurement is a rarer, but not unlikely occurrence, with a probability of $\sim16\%$.\footnote{We note that for the same dataset (September 23, 2004) as was used to compute the upper limit of 0.5~mJy, our aperture photometry method yields a flux of 1.0~mJy with aperture photometry (with S17 and the empirical background fit subtracted). This compares favourably with that of \citet[][0.8-1.0~mJy, scaled to our calibration]{sab10} obtained with aperture photometry after PSF-extraction of all known sources close to Sgr~A* including the star identified in that work as S62.} The 0.8-1.0~mJy diffuse component reported by \citet{sab10} might be associated with the $\sim$0.3-0.7~mJy offset we find in our fits to the flux distribution of Sgr~A*, between the fitted constant component (3.6~mJy) and the expected contribution of S17 (2.9-3.3~mJy). \citet{Do2009} also estimated from the color of Sgr~A* at low fluxes, that they most likely had $\sim$35\% contribution from stellar contamination (0.7~mJy) for their observations in 2006.  These results all seem to point towards an additional stellar component to the flux of Sgr~A* with a constant flux of $\sim$0.5~mJy (our data), though it does not necessarily always consist of the same stars from year to year. 

\subsection{Comparison with X-ray binary variability}
Lognormal distributions of fluxes are also seen in other astrophysical objects, for example X-ray binaries \citep[prototype Cyg X-1 in the low hard state, ][]{Uttley2005}, implying a connection between Sgr~A* other low-accretion rate black holes \cite{Falcke2004,McHardy2006}. The physical picture of the low-hard state of X-ray binaries is that -- compared to higher luminosity flows -- the inner optically thick accretion disc disappears and is replaced by a hot inner flow \citep{Done2007}. Similar pictures, with hot inner flows, exist for Sgr~A* \citep[e.g.][]{Quataert2003}. Sgr~A* is, however, at the extreme low end of the accretion rate range of observable sources (the `quiescent' state). Connections between Sgr~A* and low-hard state X-ray binaries of higher accretion rate such as Cygnus X-1 can shed light on whether or not the quiescent state is indeed a distinct state or whether there is simply a smooth continuation of the hard state down to such low accretion rates (\citealt{Markoff2010}). 

If timescale scales with mass, then the flares in Sgr~A* of $\sim$10-100~minute duration, correspond to $\approx$ millisecond timescales in a stellar mass black hole binary. It is also true that millisecond-timescale flares are seen in Cyg X-1, though it was shown by \citet{Uttley2005} that the majority of these (and all those in the low/hard state) were not more frequent than expected in the context of the extension of the low-level lognormal distribution in this source to higher fluxes. There was one flare however, a 12.4$\sigma$ event that occurred (in fact when Cyg X-1 was in the high/soft state) that could not be accounted for in the context of the underlying variability and was posited by the authors to have a different physical origin. In this vein, it could be that the flares in Sgr~A* are due to the source's ``flickering attempts at outburst activity from out of quiescence'', as suggested by \citet{Markoff2010}.

A connection between the near-infrared variability of Sgr~A* and the variability of X-ray binaries has also been made previously from a long timescale break in the PSD of Sgr~A* which can be associated with a break timescale in the low-hard state of X-ray binaries (Meyer et al. 2009), though the single dataset used to constrain the high frequency power spectrum in this case may not be representative. The lognormal distribution for low-level fluxes makes however an independent case for an analogy between the variability of Sgr~A* and low-hard state X-ray binaries.

\subsection{Wavelength Dependence}

The flux distributions of \citet{YusefZadeh2006_InfraredFlares} and \citet{yus09} at 1.6$\mu$m and 1.7$\mu$m display a qualitatively similar shape to our Ks-band distribution, although these datasets are still much smaller than ours, and the low-level activity appears to be dominated by the observational noise. Nevertheless, with more data at 1.6 and 1.7 $\mu$m and modeling of the observational errors, it might be possible to obtain constraints on the colours of the low level and high flux components.  
In future work we intend to use the same techniques as we have used in this paper at Ks-band, in other bands (L' and H) to investigate in detail the dependence of the flux distribution with wavelength.

\subsection{Timing Analysis}

Our work is not the first to point out that the flux distribution of Sgr~A* is not well-described by a Gaussian: \citet{YusefZadeh2006_InfraredFlares}, \citet{Do2009} and \citet{yus09} have all noted that the flux distribution has a high flux tail. However, the timing analyses which have been carried out to date \citep{Do2009,Meyer2008,Meyer2009} in which Sgr~A* was compared to simulated red noise lightcurves have nevertheless used simulated lightcurves which have had, by construction, a Gaussian flux distribution. Sgr~A*'s intrinsic mean and variance are such that if one were to plot the simulated lightcurves scaled to Sgr~A*'s mean and variance (for the 2004-2009 dataset $\mu=1.3$~mJy subtracting 3.8~mJy for S17 plus other faint stars, and $\sigma=1.9$~mJy), the model red noise lightcurves would have negative intrinsic fluxes much of the time which is rather unphysical. 

In \citet{Do2009} the high flux tail was excluded when fitting the flux distribution, and only fluxes $\lesssim$3.2~mJy were fitted with a Gaussian (see Table \ref{table2} in the Appendix for scaling factors; the reported value in \citealt{Do2009} was a reddened flux of 0.3~mJy). Thus their model lightcurves best resemble the low level emission of Sgr~A*. Indeed the entire dataset used by \citet{Do2009} does not include much bright emission $\gtrsim 5$~mJy, and thus their results really apply to the low level emission of Sgr~A*. That the low level emission from Sgr~A* is consistent with red noise fits together with our finding of the lognormal distribution of fluxes at low fluxes. Both of these aspects of the variability are seen in X-ray binaries such as Cyg X-1 \citep{Uttley2005}.  
However, given that the flux distribution is not Gaussian, it remains questionable that the power spectrum has yet been reliably constrained from the Monte Carlo simulations of previous timing studies. 

In particular, high and low flux states of Sgr A* may require separate timing analysis. In this respect it is interesting that \citet{Meyer2008} finds a quasi-periodic signal with false alarm probability of only $2\times10^{-5}$ (4.2$\sigma$ significance in Gaussian equivalent terms), if one considers just the window 385-445~minute subset from 30-31 July 2005, though the signal was deemed insignificant when the entire lightcurve from the night was analyzed. 

\section{Conclusions}\label{section_conclusions}

We have determined the distribution of fluxes for the variable source Sgr~A* in the near-infrared and thereby investigated flux-dependent properties of the emission. We summarize our main results as follows:
\begin{itemize}
 \item Sgr~A* is continuously emitting and variable in the near-infrared.
 \item At low Ks-band fluxes the variability follows a lognormal distribution with a median flux of 1.1 mJy and a multiplicative standard deviation factor of 2.1. We identify this continuously variable low-level state as the quiescent state of Sgr~A* in the near-infrared.
 \item At least 0.5 mJy of the flux of Sgr A* is due to a faint stellar component at all times during 2004-2009.
 \item There is variability on a $\approx$two-week timescale in the low level component, which can not be attributed to stars.
 \item At high fluxes (above $\sim$5~mJy) the flux distribution flattens which seems to be most likely explained by the presence of sporadic flare events additional to the quiescent emission. 
 \item On August 5th 2008 we observed a very bright Ks-band flare of 27.5~mJy, the brightest Ks-band flare yet observed from Sgr~A*. 
\end{itemize}
Differences in spectral index, polarization properties and association with X-ray flares may already be further indications of the different nature of the high flux near-infrared emission from Sgr~A*. In future work we aim to search for and quantify further possible differences between high and low flux emission of Sgr~A*.

\end{document}